Optical Extinctions of Inter-Arm Molecular Clouds in M31: A Pilot Study for the Upcoming

*CSST* Observations

Cailing Chen[1,2], Zheng Zheng[1,3], Chao-Wei Tsai[1,3,4], Sihan Jiao[1,3], Jing Tang[1], Jingwen Wu[2,1,3], Di Li[1,2,3,5], Yun Zheng[5], Linjing Feng[1,2], Yujiao Yang[1], Yuan Liang[2]

[1] National Astronomical Observatories, Chinese Academy of Sciences, Beijing 100101, China

  *zz@bao.ac.cn, cwtsai@nao.cas.cn, sihanjiao@nao.cas.cn*

[2] University of Chinese Academy of Sciences, Beijing 100049, China

[3] Key Laboratory of Radio Astronomy and Technolgoy, Chinese Academy ofSciences, Beijing 100101, China

[4] Institute for Frontiers in Astronomy and Astrophysics, Beijing Normal University, Beijing 102206, China

[5] Zhejiang Lab, Hangzhou, 311121, China

**Abstract** Recent submillimeter dust thermal emission observations have unveiled a significant number of inter-arm massive molecular clouds in M31. However, the effectiveness of this technique is limited to its sensitivity, making it challenging to study more distant galaxies. This study introduces an alternative approach, utilizing optical extinctions derived from space-based telescopes, with a focus on the forthcoming China Space Station Telescope (*CSST*). We first demonstrate the capability of this method by constructing dust extinction maps for 17 inter-arm massive molecular clouds in M31 using the Panchromatic Hubble Andromeda Treasury (PHAT) data. Our analysis reveals that inter-arm massive molecular clouds with an optical extinction ($A_V$) greater than 1.6 mag exhibit a notable $A_V$ excess, facilitating their identification. The majority of these inter-arm massive molecular clouds show an $A_V$ around 1 mag, aligning with measurements from our *JCMT* data. Further validation using a mock *CSST* RGB star catalog confirms the method's effectiveness. We show that the derived $A_V$ values using *CSST* $z$ and $y$ photometries align more closely with the input values. Molecular clouds with $A_V >1.6$ mag can also be identified using the *CSST* mock data. We thus claim that future *CSST* observation could provide an effective way for the detection of inter-arm massive molecular clouds with significant optical extinction in nearby galaxies.

# 1 INTRODUCTION

Molecular clouds, especially massive molecular clouds, serve as nurseries of young stars and play a crucial role in star formation and galaxy evolution (Dame et al. 1987; Solomon et al. 1987; Gao & Solomon 2004; Leroy et al. 2008, 2013; Chevance et al. 2023). In spiral galaxies, molecular clouds tend to distribute along the spiral arms. For instance, in M31, the two prominent spiral arms and a star-forming ring at 10 kpc contain significant amounts of dust and gas, resulting in the highest star formation rates within M31 (Habing et al. 1984; Rice 1993, Hu & Peng 2014).

In addition to the extensively studied molecular clouds within the spiral arms of galaxies, there are also molecular clouds residing in the inter-arm regions (Koda et al. 2009; Colombo et al. 2014; Leroy et al. 2021). These inter-arm clouds exhibit lower column densities, which result in comparatively weaker star formation activities when contrasted with their counterparts in the spiral arms. Due to the edge-on projection of the Milky Way and the challenges associated with line-of-sight distances, obtaining a comprehensive view of the molecular clouds within the Milky Way is an arduous task. Identifying and studying molecular clouds located between the arms is particularly challenging. However, we have made a significant discovery by identifying a considerable number of compact molecular clouds (< 54 pc at the distance of M31) in both on-arm and inter-arm regions (Jiao et al. in prep.). This was made possible through exceptionally deep 850 $\mu$m continuum mapping observations using the *JCMT-SCUBA2*. The inter-arm clouds in M31 consistently exhibit lower masses than typical clouds on arms, some are confirmed by the following up IRAM 30m and NOEMA CO 1-0 observations. The unique characteristics of these massive inter-arm molecular clouds render them particularly intriguing subjects for further systematic and detailed studies using a larger sample. However, carrying out a systematic survey for inter-arm molecular clouds in galaxies beyond M31 using single-dish submillimeter telescopes would be nearly impossible due to resolution and confusion limitations, and using interferometers like the Atacama Large Millimeter/submillimeter Array (ALMA) would be too expensive.

A viable alternative for studying these inter-arm molecular clouds is through optical extinctions. The optical extinction method, with its long-standing history in dust measurement dating back to the work of Bok et al. (1937) and Bok (1956), provides a crucial and straightforward method for observing molecular clouds (Lombardi & Alves 2001). Most previous research on optical extinctions is confined to the Milky Way and its close neighbors such as the Magellanic Clouds (e.g. Dickman 1978; Lada et al. 1994; Lombardi & Alves 2001) due to lack of high spatial resolution multi-band large-area observations. The forthcoming China Space Station Telescope (*CSST*), with its subarcsecond resolution and multi-band coverage and extensive survey area, could potentially offer a novel approach to systematically detecting inter-arm molecular clouds in nearby galaxies.

Indeed, there have been a few studies on dust extinctions using observations from space telescopes. For example, the Panchromatic Hubble Andromeda Treasury (PHAT) survey, conducted by Dalcanton et al. (2012), targeted M31 and provided extensive coverage of one-third of the M31 disk with multi-band and high spatial resolution. This survey offers an opportunity for studying dust properties across various scales, from galactic to stellar. Dalcanton et al. (2015) derived an optical extinction map of the northeast part of the M31 disk using the Color-Magnitude Diagram (CMD) of red giant stars(Williams et al. 2014). Lindberg et al. (2024) studied dust properties around massive stars through the fitting of multi-band Spectral Energy Distributions (SEDs) of individual stars. While these studies leverage the PHAT survey data effectively to explore optical extinctions and dust properties, they have not specifically addressed the extinctions associated with inter-arm molecular clouds.

We thus propose to explore the possibility of using *CSST* data to study the inter-arm molecular clouds by calculating optical extinctions, aiming to systematically study these components in nearby galaxies. We first utilize PHAT data to demonstrate the feasibility of such a study using a subset of inter-arm molecular clouds in M31. We then extend the methodology to *CSST* mock data. Once validated, *CSST* data will greatly facilitate the exploration of the optical properties of extra-galactic molecular clouds. The structure of our paper is laid out in the following manner: Section 2 provides a concise overview of the massive molecular clouds sample, the PHAT data and the CMD methodology (Dalcanton et al. 2015). Section 3 presents the calculation results using PHAT data and discusses our findings. In Section 4, we briefly introduce the *CSST* parameters and apply our method to *CSST* mock data. We summarize our work and present our conclusions in Section 5.

## 2 SAMPLE, DATA AND METHODOLOGY

As previously mentioned, Dalcanton et al. (2015) have produced an overall dust optical extinction map for the northeastern part of the M31 disk utilizing the PHAT dataset. Given this groundwork, it is a logical step to utilize their data and methodology as a starting point for our preliminary analysis. Accordingly, we select a subset of inter-arm massive molecular clouds from our extensive *JCMT* surveys (Jiao et al. in prep., Jiao et al. 2022) that overlap with the PHAT coverage. We apply the methodology established by Dalcanton et al. (2015), incorporating minor adjustments to the spatial grid parameters to better reflect the sizes of the inter-arm massive molecular clouds. We briefly outline the inter-arm massive molecular cloud samples, the data and the methodology. A more extensive discussion of our *JCMT* survey and the properties of the inter-arm massive molecular clouds will be presented in a forthcoming paper by Jiao et al. (in prep.). For in-depth descriptions of the PHAT data and methodology, readers are referred to Dalcanton et al. (2012, 2015).

These clouds were identified and characterized based on the combined continuum image at the 850 *μm* band. The subsample contains 17 inter-arm massive clouds, which are located in the PHAT footprint. We first examine the optical images of these massive molecular clouds and their surroundings, hoping to identify dark regions akin to those cataloged by Lynds (1962). We show 3-color images constructed using the F160W, F110W, and F814W filters in Fig. 1. Based on a visual inspection of the three-color images, we notice that the regions corresponding to the massive molecular clouds (black circles in Fig. 1) do not seem to contain any pronounced dark clouds that would differentiate them from other areas within the images. The reason for this may be due to the relatively low level of dust extinction in these areas, which makes it barely visible in the images. Quantitative analysis of $A_V$ using the CMD method is thus necessary to gain a better understanding of the dust extinction in these regions.

The PHAT survey covers the north-eastern part of the M31 star-forming disk using *HST* Advanced Camera for Surveys (ACS), Wide Field Channel (WFC), and the Wide Field Camera 3 (WFC3) in six bands from the near ultraviolet (F275W, F336W), optical (F475W, F814W), to the near-infrared (F110W, F160W). The survey footprint (see also Fig. 3) and detailed parameters were described in Dalcanton et al. (2012). Williams et al. (2014) has measured photometries of 117 million resolved stars across all the six bands. The limiting magnitude in the most crowded regions reaches F475W ~25 mag and is enough for analysis using RGB stars. We thus take advantage of the stellar catalog by Williams et al. (2014), and conduct our study by selecting individual red giant branch (RGB) stars in F110W and F160W filters. We exclude the 'brick' located at the center of M31 due to its high stellar crowding, which hinders our analysis. We follow the data cleaning procedure described in Dalcanton et al. (2015) to filter out the necessary data for further analysis.

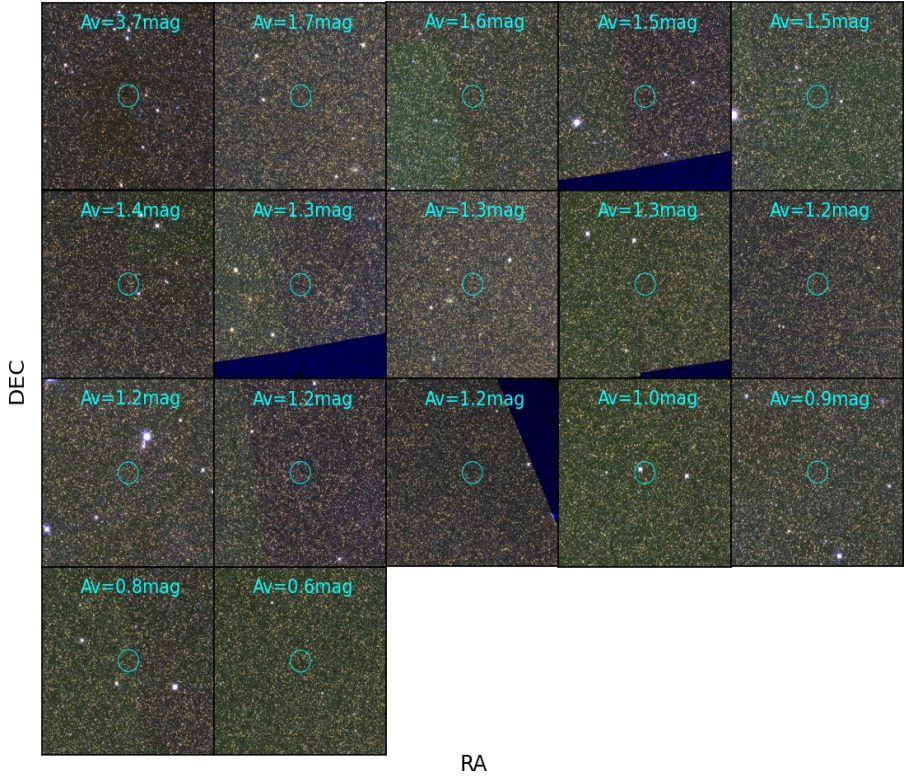

Fig. 1: Three-color images of all 17 massive molecular clouds and their surroundings, captured in F160W (red), F110W (green), and F814W (blue) bands. The images are arranged in descending order based on their extinction value (calculated in this paper). The size of each panel is $2'\times2'$ and the circles have a diameter of $14''$ (corresponding to the resolution of Jiao et al.(in prep.)).

We implement the approach proposed by Dalcanton et al. (2015), with the modification of utilizing slightly adjusted gridding parameters for the computation of the $A_V$ value. The fundamental assumption of Dalcanton et al. (2015) is that the dust is distributed in a thin layer embedded within a thick stellar disk. Consequently, the light from RGB stars situated behind the dust layer is subjected to extinction, while the light from the remaining RGB stars positioned in front of the dust layer remains unaffected by dust extinction. Theoretically, the distribution of RGB stars without dust extinction in the NIR CMD should form a nearly vertical narrow sequence. This vertical sequence would broaden and shift toward a redder color if obscured by the dust layer. Consequently, the CMD would exhibit two sequences (see Fig. 2 for examples): the broad red sequence is attributed to the RGB stars located behind the dust layer, and the narrow blue sequence is attributed to the RGB stars located in front of the dust layer. In this study, we assume that all the inter-arm massive molecular clouds are distributed within this thin layer due to their low scale height compared to the thick stellar disk (Stark & Lee 2005; Jefferson et al. 2022). Then we could conduct fitting procedures to determine the extent of displacement between the unreddened and reddened sequences and thereby obtain the extinction value. Following Dalcanton et al. (2015), we select regions with similar stellar densities in the outer disks to generate reference CMD models with zero $A_V$. We apply Cardelli et al. (1989) extinction law with $A_{F110W}/A_V = 0.3266$, $A_{F160W}/A_V = 0.2029$, and a standard $R_V = 3.1$.

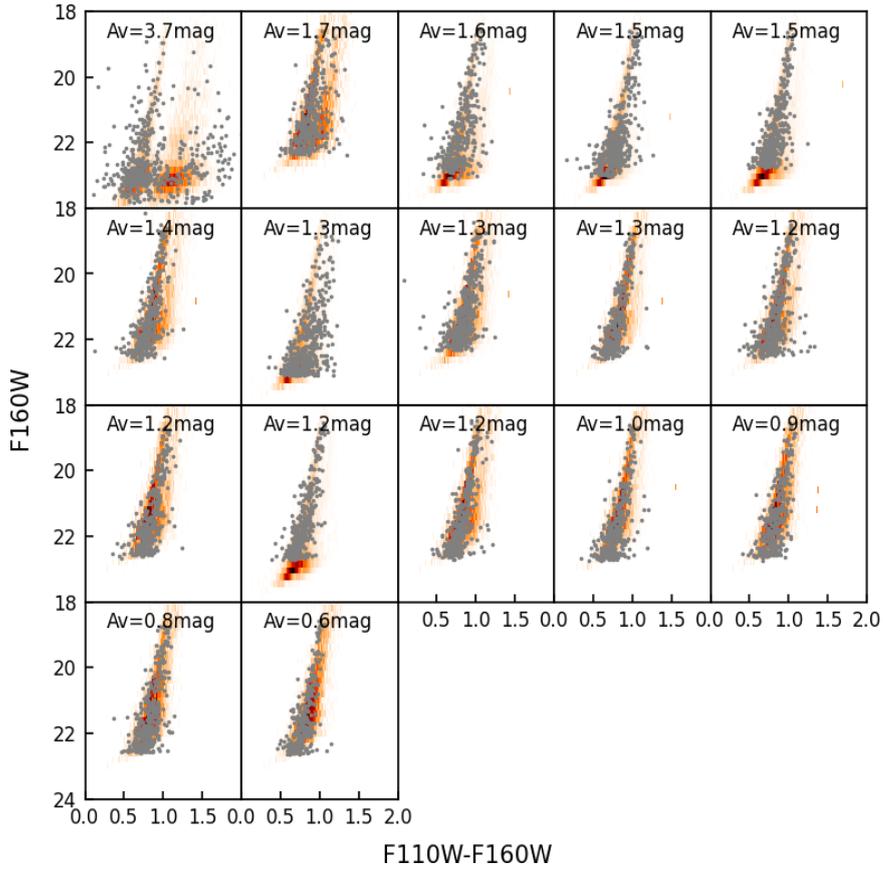

Fig. 2: The F110W-F160W color versus F160W magnitude diagrams (CMDs) of RGB stars in 17 inter-arm massive molecular cloud regions (at the circled locations in Fig. 1), with sizes of $14''$ bins. The brown stripes in the background show the model fits. The panels are arranged in descending order by their extinction value. CMDs with large $A_V$ exhibit more reddish points and a larger color dispersion.

Initially, RGB stars are selected from the Williams et al. (2014) catalog within a $7'' \times 7''$ ($\sim$26 pc at the distance of M31) bin centered around our target location to gain a better angular resolution. But for the convenience of comparing with emission-based $A_V$ by Jiao et al.(in prep.), we specifically calculate the the selected massive molecular clouds with a diameter of $14''$. Subsequently, we perform Bayesian fitting on the F110W-F160W v.s. F160W CMD of the RGB stars within the square bin using a set of four parameters: the median optical extinction $\tilde{A}_V$, the dimensionless width of the extinction $\sigma$, the acceptable range of color shift of the unreddened CMD (a few hundredths of a magnitude) $\delta_c$, and the fraction of stars behind the dust layer $f_{red}$. It is noteworthy that we apply a uniform prior distribution to the $\tilde{A}_V$ parameter. The lower bound of this distribution was set at 0, while the upper bound is determined based on the observed reddening of RGB stars relative to unreddened RGB stars in the CMD. Additionally, the determination of this upper bound is informed by referencing the published emission-based $A_V$ value from Draine et al. (2014). The fitting is executed using the *emcee* software (Foreman-Mackey et al. 2013) which employs the Markov Chain Monte Carlo (MCMC) method. The MCMC sampler of grids are run in batches defined by the survey area.

## 3 RESULTS

We first present the locations of our inter-arm and a sample of on-arm massive molecular clouds from Kirk et al. (2015) on the full map of dust extinction in the PHAT footprint (Fig. 3). The inter-arm massive molecular clouds are generally located in the low optical extinction regions ($A_V \sim 0.5$) wile the on-arm massive molecular clouds are located in high optical extinction environments ($A_V \sim 2$). Their $A_V$ values are calculated and presented in Table 1. It also provides detailed information on these inter-arm clouds, encompassing the values of CMD-based $A_V$ and emission-based $A_V$ as reported in Draine et al. (2014) and Jiao et al.(in prep.).

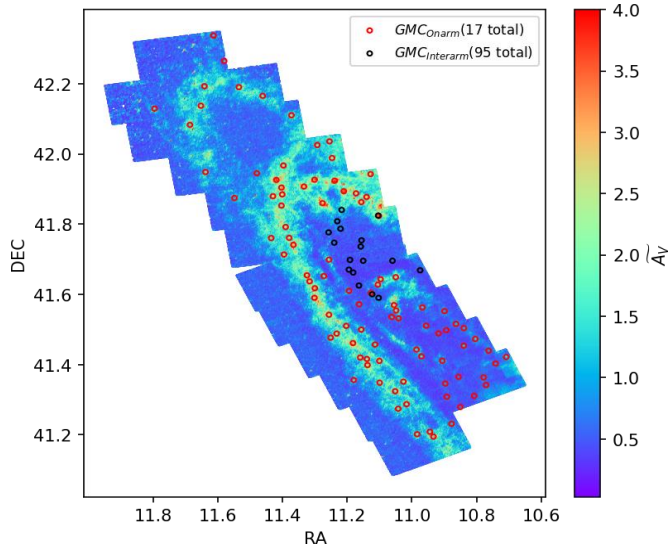

Fig. 3: Map of the median extinction $\tilde{A}_V$ value for the whole PHAT footprint except for the most central regions of M31. We show the inter-arm and on-arm massive molecular clouds using black and red circles, respectively.

We present the distribution maps of $\tilde{A}_V$ around the targeted inter-arm massive molecular clouds in Fig. 4. The black circles, each with a diameter of $14''$, mark the location of each massive molecular cloud. Notably, for some clouds with $\tilde{A}_V > \sim 1.6$ mag (which corresponds to a $2\sigma$ deviation), their $\tilde{A}_V$ values are significantly higher than the surrounding regions, highlighting the efficiency of our extinction analysis in capturing such variations. However, for the majority of the massive molecular clouds, there are no sub-stantial differences in $\tilde{A}_V$ values compared to the surrounding regions. The main reason for this could be the relatively

Table 1: Extinction value of massive molecular clouds

| Label | RA | DEC | $\widetilde{A_V}$ | $\Delta \widetilde{A}_V$ | $A_{V,Hershel}$ | $A_{V,JCMT}$ |
|-------|-----|------|-------|--------|-------|--------|
| 1 | 0h44m24.91s | 41°49′28.13″ | 3.7 | 0.8 | 8.8 | 9.7 |
| 2 | 0h44m29.42s | 41°36′01.31″ | 1.7 | 0.7 | 2.3 | 3.4 |
| 3 | 0h45m01.81s | 41°46′36.35″ | 1.6 | 0.8 | 1.5 | 2.7 |
| 4 | 0h44m37.92s | 41°44′10.93″ | 1.5 | 0.7 | 0.9 | 2.3 |
| 5 | 0h44m57.59s | 41°44′49.93″ | 1.5 | 0.9 | 1.2 | 3.5 |
| 6 | 0h44m14.36s | 41°41′45.20″ | 1.4 | 0.8 | 1.6 | 4.3 |
| 7 | 0h44m52.07s | 41°50′26.88″ | 1.3 | 0.8 | 2.8 | 4.3 |
| 8 | 0h44m24.66s | 41°35′24.84″ | 1.3 | 0.9 | 1.3 | 3.5 |
| 9 | 0h44m43.48s | 41°39′44.01″ | 1.3 | 0.8 | 0.8 | 2.9 |
| 10 | 0h44m39.26s | 41°37′30.88″ | 1.2 | 1.0 | 0.9 | 2.0 |
| 11 | 0h44m55.30s | 41°48′30.05″ | 1.2 | 0.9 | 1.8 | 3.1 |
| 12 | 0h44m37.35s | 41°45′14.29″ | 1.2 | 0.9 | 1.1 | 2.6 |
| 13 | 0h43m53.81s | 41°40′06.87″ | 1.2 | 0.8 | 1.1 | 4.1 |
| 14 | 0h44m46.76s | 41°40′13.85″ | 1.0 | 0.8 | 0.8 | 2.4 |
| 15 | 0h44m52.89s | 41°47′13.51″ | 0.9 | 0.7 | 0.9 | 3.0 |
| 16 | 0h44m36.08s | 41°41′44.35″ | 0.8 | 1.0 | 1.0 | 5.0 |
| 17 | 0h44m45.91s | 41°41′53.89″ | 0.6 | 1.2 | 0.9 | 2.6 |

Notes: This table presents the $A_V$ value of interarm massive molecular clouds in different researches. Column (1) massive molec- ular cloud label, (2) R.A. (RA), (3) decl. (DEC), (4) $\tilde{A}_V$ value computed in this paper, (5) $\tilde{A}_V$ uncertainty from this paper (the mean uncertainty $\sigma$ is ∼0.7 mag), (6) emission-based $A_V$ from Draine et al. (2014), (7) emission-based $A_V$ from Jiao et al. (in prep.).

large uncertainty ( ∼0.7 mag)and relatively low optical extinction of the inter-arm massive molecular clouds.

Another possible speculation is that the RGB samples we choose for $A_V$ calculation are mostly located in front of these inter-arm massive molecular clouds, which may lead to an underestimation of the actual $A_V$ value. We thus compare the number of RGB stars in the massive molecular cloud (both inter-arm and on-arm) regions and that in other regions without molecular clouds (see Fig. 5). It shows that the distributions of RGB star counts are similar in all three types of regions. We thus suggest that this speculation is not preferable.

While the CMD method may not be effective for identifying inter-arm massive molecular clouds with low optical extinction ($A_V < 1.6$ mag), it does demonstrate that optical/infrared imaging data from space telescopes such as *HST* can serve as a viable alternative to submillimeter observations in detecting massive molecular clouds with higher extinction ($A_V > 1.6$ mag) in nearby galaxies. In the region covered by PHAT survey in M31 (approximately one-third of M31 disk), 17 inter-arm massive molecular clouds were detected by *JCMT* through long time integration (Jiao et al. in prep.). Among them, three clouds are found to have an $A_V$ greater than 1.6 mag using the CMD method, making them detectable in optical surveys. The fraction of such high-$A_V$ clouds in the PHAT footprint is roughly 18%, which is consistent with the findings of Jiao et al. (in prep.) who reported a similar percentage of 20% for high-$A_V$ clouds in a larger area. Although the CMD method only reveals a small fraction of these inter-arm massive molecular clouds, these heavily extincted clouds tend to be the most massive molecular clouds between the spiral arms, and they are crucial for studying the formation and evolution of inter-arm molecular clouds

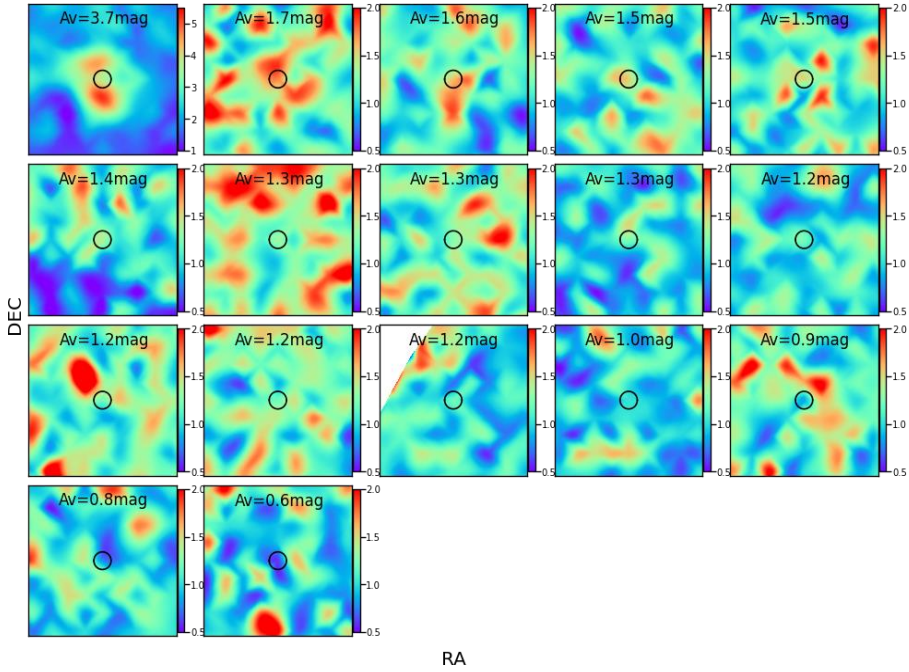

Fig. 4: The interpolated and smoothed $A_V$ distribution maps of all massive molecular clouds, as well as of their respective surrounding regions (total $2' \times 2'$). Three of them (indicated by black circles with a diameter 14″) with $A_V \gtrsim 1.6$ mag exhibit an excess of extinction values compared to their surroundings. Massive molecular clouds with lower $A_V$ values cannot be identified by eye in these $A_V$ maps.

Utilizing wide-field optical surveys, we can identify them independently of deep dust integration, making this approach highly valuable.

Following the analysis by Dalcanton et al. (2015), we plot the extinction values of all the bins in the PHAT footprint (small gray dots) derived by our CMD-based method versus those derived from 3.6-500 $\mu m$ dust emissions in Draine et al. (2014) in Fig. 6. The distribution of the gray dots is generally consistent with the results of Dalcanton et al. (2015). In the low extinction regions ($A_V < 1$), no discernible correlation is observed with the value of Draine et al. (2014). In the high extinction regions ($A_V > 1$), a proportional correlation of approximately 2.5 times to the $A_V$ values from Draine et al. (2014) is established (black solid line in Fig. 6; see Dalcanton et al. (2015) for more discussion).

We mark both inter-arm massive molecular clouds (green filled circles; correspond to the black circles in Fig. 3) and normal on-arm massive molecular clouds (red dots; correspond to the red circles in Fig. 3) in Fig. 6. The inter-arm massive molecular clouds have optical extinction value $A_V \sim 1$ and the on-arm massive molecular clouds have a much wider $A_V$ distribution across 0.5 to 4. It is noteworthy that all the green dots are located under the black solid line that represents the systematic ratio of 2.5 between dust emission-obtained $A_V$ and CMD-obtained $A_V$ in Dalcanton et al. (2015) , while the normal on-arm massive molecular clouds generally follow the distribution of the gray dots. The systematic discrepancy has been observed in Dalcanton et al. (2015) and a few references therein (Wang et al. 2022). This could be due to various reasons such as different methods and distinct resolutions of dust. Further discussion on this topic is out of the scope of this paper. The deviation of the inter-arm massive molecular clouds from the systematic 2.5-offset suggests that the inter-arm massive molecular clouds may have a lower far-infrared/submillimeter emission than the bulk of dust clouds with similar optical $A_V$.

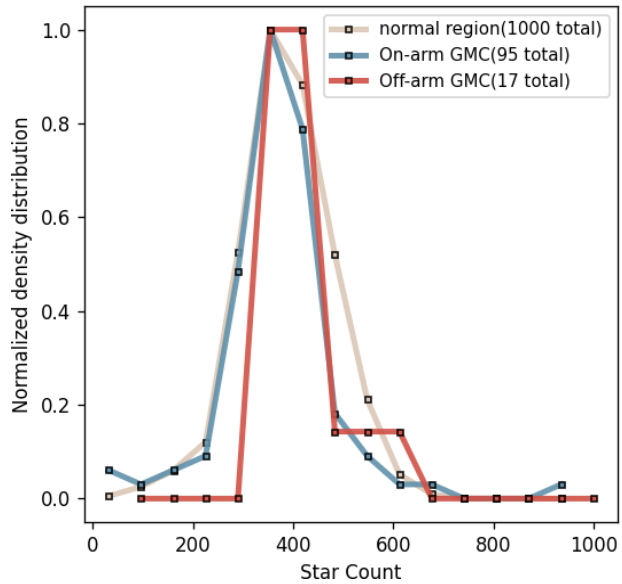

Fig. 5: RGB Star counts vs. their corresponding probability density, i.e., the star count distributions of on-arm, inter-arm massive molecular clouds as well as 1,000 normal regions in M31. The 1,000 normal regions are randomly chosen in areas away from the molecular clouds. The star counts are consistent across these three types of regions, indicating that the number of RGB stars not visible due to high extinction is very low.

We also overplot the extinction value derived from Jiao et al.(in prep.), $A_{V,JCMT}$, which are based on observations with dust emissions including a longer wavelength (850 $\mu m$), in Fig. 6 (filled orange squares). The $A_{V,JCMT}$ generally has a value 1-3 magnitudes higher than that derived by Draine et al. (2014), aligning more closely with the gray dots and black solid line in the figure. This suggests that inter-arm massive molecular clouds exhibit higher emissions at longer wavelengths, indicating a lower temperature compared to their counterparts in the spiral arms (a comprehensive discussion on this will be presented in Jiao et al. (in prep.)). The agreement between $A_{V,CMD}$ and $A_{V,JCMT}$ further supports the superiority of the CMD method in studying inter-arm massive molecular clouds over dust emission surveys that focus on shorter wavelengths.

## 4 FUTURE PERSPECTIVES AND TEST USING THE *CSST* MOCK DATA

As shown in the previous section, optical/infrared imaging data from space telescopes offers a viable way to detect high extinction ($A_V$>1.6 mag) inter-arm massive molecular clouds in nearby galaxies. Recently, there have been a few more observations similar to PHAT. For the southern part of M31, the Panchromatic Hubble Andromeda Southern Treasury (PHAST, Williams et al. 2021a) survey has just been completed. The data are expected to be released in 2024, and then the dust map of the entire M31 will be available. In M33, a similar survey called the Panchromatic Hubble Andromeda Treasury Triangulum Extended Region (PHATTER, Williams et al. 2021b) was carried out, where this method can also be used. Recently, high resolution observations from several telescopes, including *ALMA*, *HST*, *JWST* and *VLT*, are combined to study 74 nearby spiral galaxies within 20 Mpc by the program PHANGS (Physics at High Angular resolution in Nearby Galaxies; Lee et al. 2022, 2023). Among them, 55 galaxies will have images in *HST/WFC3/UVIS* F814W and *JWST/NIRCam* F150W. It is worth investigating whether this method is applicable to these data in the future.

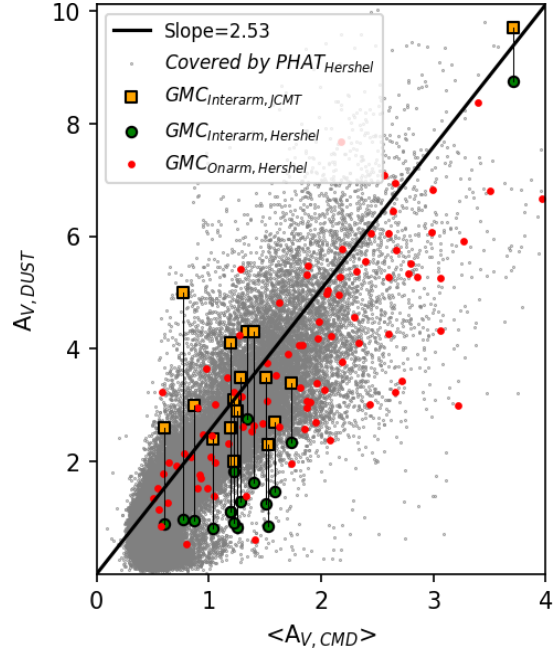

Fig. 6: Pixel-by-pixel comparison of the dust emission-based $A_V$ (y-axis) and the CMD-based mean $A_V$ (x-axis). The gray points show all the pixels in Fig. 3. The gray points follow a ratio of $A_{V,DUST}/A_{V,CMD}$ ~2.5, as indicated by the thick black solid line. The $A_{V,DUST}$ is the extinction es- timated from the dust emission at $3.6 - 500$ micron when Draine et al. (2014) is adopted, or at 850 micron when Jiao et al.(in prep.) is used. The $A_{V,CMD}$ indicates the extinction derived from the CMD method from this paper. The orange squares correspond to the $A_V$ values of inter-arm massive molecular clouds derived by Jiao et al.(in prep.), which includes 850 $\mu m$ dust emissions; and the green dots correspond to the $A_V$ values of these inter-arm massive molecular clouds derived by Draine et al. (2014), which uses mostly shorter wavelength dust emissions. The values of the same cloud are connected by a thin line. We also mark the on-arm massive molecular clouds from Draine et al. (2014) by red dots. The green dots are systematically lower than the thick black line, while the orange squares and red dots do not show much deviation from the thick black line.

Apart from these surveys, the upcoming *CSST* will conduct a wide-field multiband imaging survey in *NUV*, *u*, *g*, *r*, *i*, *z* and *y* bands, covering the wavelength range of 0.255-1 $\mu$m. *CSST* is a space telescope with a diameter of 2 meters. Its field of view is 1.1 deg$^2$ and spatial resolution is ~0.15``. The survey is designed to cover 17500 deg$^2$ with an exposure time of 150×2 s for each band. The point-source 5$\sigma$ depth in *g* and *r* bands can reach 26 AB mag. In addition, there will be 400 deg$^2$ selected as deep fields to observe more time (250×8 s for each band) to reach one-magnitude deeper. More details can be found in Zhan (2021, 2011).

Different from target surveys conducted by *HST* and *JWST*, *CSST* survey will provide much more complete sample of nearby galaxies with various types. Furthermore, photometric bands with shorter wavelengths, such as the NUV/u filters, are more sensitive to dust absorptions, which results in larger reddening on the CMD even with minimal dust extinction. However, shorter wavelengths also present challenges. According to Dalcanton et al. (2015), using data from filters with shorter wavelengths, such as F475W and F814W, may introduce large uncertainties. Thus it makes sense to validate the CMD method for *CSST* data to ensure its reliability.

To achieve this goal, we generate *CSST* mock catalog of RGBs in M31 to test this method. We select a number of 'unreddened' RGBs, that is, without extinction by interstellar dust, in PHAT data according to Dalcanton et al. (2015), and correct the data for the Galactic extinction of $A_V = 0.17$ (Schlafly & Finkbeiner 2011), following the CCM extinction law (Cardelli et al. 1989). We fit the intrinsic photometry of these individual RGBs using grids of stellar atmosphere models with a range of $T_{\text{eff}}$, gravity and metallicity, and convert them to *CSST* photometric magnitudes, based on Chen et al. (2019). We then take into account the Galactic extinction, and add the photometric errors as follows:

$$\text{星等误差} = 2.5ln10 \times \frac{f_{err}}{f} = 2.5ln10 \times \frac{1}{SNR}$$

where $f$ and $f_{err}$ represent the flux of signal and noise respectively. The signal-to-noise ratio (*SNR*) is calculated using the exposure time calculator[1] of *CSST* with specific bands and typical exposure times. Note that the noise here only considers photon shot noise from astronomical objects and sky background, as well as instrumental noise including dark current noise and readout noise (Zhang et al. 2019). The error induced by data reduction is not taken into account. Next we test the CMD method in both non-cloud and cloud regions.

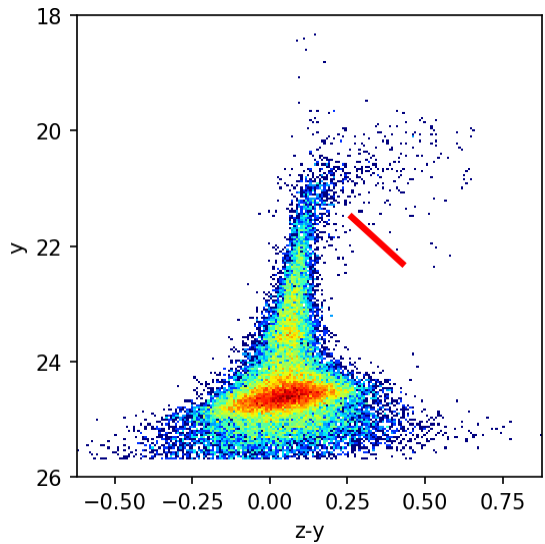

Fig. 7: Hess diagram of our unreddeded *CSST* mock data. The red bar indicates the reddening vector for $\Delta A_V = 2$. The distribution shows that the width of RGB is thinner compared to the reddening vector.

We first present in Fig. 7 the Hess diagram of *CSST* mock data ($A_V=0$) and compare its dispersion with the reddening vector (the red bar) when $\Delta A_V$ is set to 2. Comparing this with the Hess diagram for F110W and F160W filters depicted in Dalcanton et al. (2015) Figure 38, the differences are minimal and the width of the RGB is sufficiently thin in both filter combinations. Consequently, the *CSST z* and *y* band data could, in theory, be suitable for studying massive molecular clouds with large extinctions.

---

[1] https://nadc.china-vo.org/csst-bp/etc-ms/etc.jsp

We then apply artificial $A_V$ value using the extinction law (Cardelli et al. 1989) into our *CSST* mock data and calculate $A_V$ value using the RGB population splitting technique. We assess this method based on the ability to recover the input $A_V$ value. Specifically, we divide the mock data into $7''\times7''$ bins and employ extinction models with log-normal distribution. These models have different mean values (i.e., 0.5, 2, and 4) and standard deviations (0.1 and 0.3) for each bin. A total of four filters (i.e., *r*, *i*, *z* and *y*) from *CSST* are utilized and pairwise combinations are tested for all the filter combinations. The results, shown in Fig. 8, reveal that the combination of the two longest wavelength bands (i.e., *z* and *y* band) is best to recover the input extinction value accurately. For other combinations, their outcomes deteriorate with a shift toward shorter wavelengths.

Fig. 8 also presents results of four different signal-to-noise ratios (dependent on exposure time) in four filter combinations. The horizontal axis represents the input $A_V$ value, and the vertical axis represents substraction of input $A_V$ from calculated $A_V$ by CMD fitting. In our testing, the capability to recover input $A_V$ is enhanced with increasing exposure time and a redder filter, reaching the optimal value in the *z* and *y* filter combination. In Conclusion, our method accurately recovers the input extinction values with acceptable uncertainties ($\sim 0.5mag$) in the *z*, *y* band combination, demonstrating its feasibility and reliability.

We further evaluate the capability of the method in detecting inter-arm massive molecular clouds. We simulate a sample of RGB stars and distribute them evenly across a grid comprising 121 bins. Each bin corresponds to a $14''\times 14''$ area. One of the 121 bins is chosen as the massive molecular cloud location, while the others are considered normal regions.

We then apply input extinction values across the grid and calculate the resulting extinctions for each bin using our method. For this purpose, we choose the combination of *z* and *y* bands, using an exposure strategy of 8×250s. A total of six experiments were performed, each with a mean extinction value of 0.8, 1.5, 2.0, 3.0, 4.0, and 5.0 mag in the cloud bin. We assume the extinction values has a log-normal distribution with a standard deviation of 0.3 mag. For the bins without a massive molecular cloud, the mean input extinction is 0.5 mag with a standard deviation of 0.3 mag. These values were determined based on the fluctuation of extinction values among the normal regions in the PHAT data. The derived extinction maps are shown in Fig. 9. The maps indicate that our method can successfully identify massive molecular clouds with elevated ($A_V$>1.5 mag), consistent with our findings using the PHAT data.

Finally, we must emphasize that despite our tests have yielded promising results, we expect to face a range of challenges when working with future real *CSST* data. For example, given that the NIR data captured by the *HST* primarily features RGB stars, the task of selecting RGB samples poses no concern when utilizing *HST* data. However, since the *CSST* data encompasses bluer bands, utmost caution must be exercised in carefully sieving through the *CSST* data to identify appropriate RGB samples.

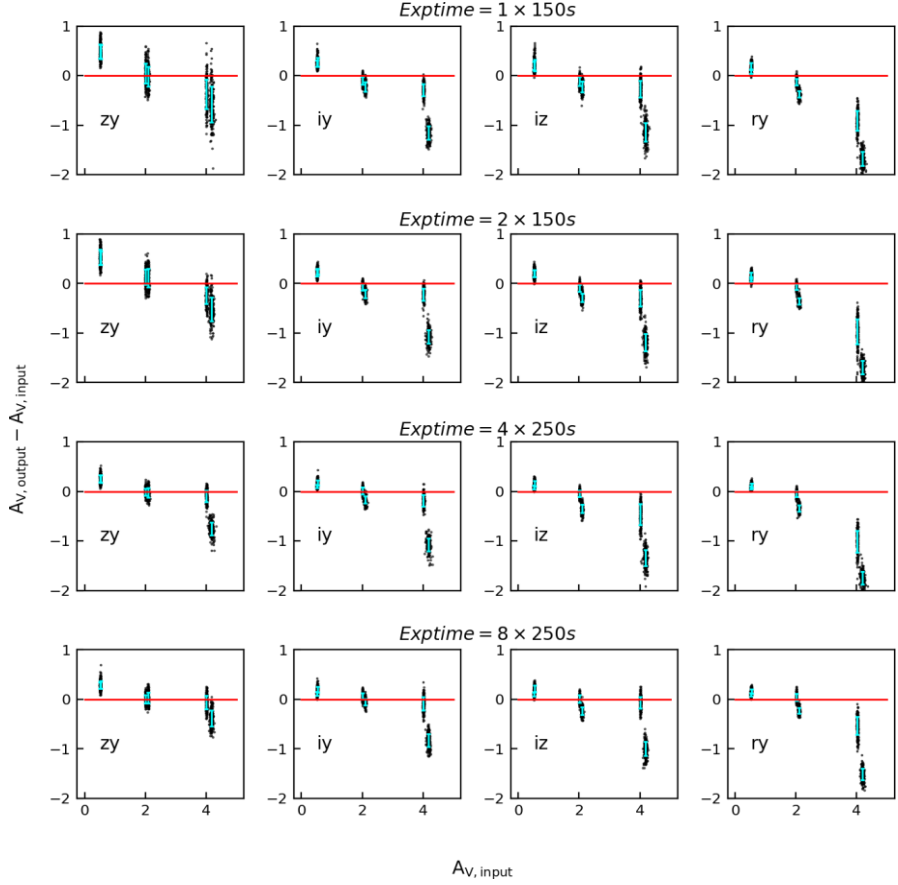

Fig. 8: $A_{V,input}$ vs. $A_{V,output} - A_{V,input}$. Panels from top to bottom represent different exposure times (which correspond to the varying signal-to-noise ratios). The panels arranged from left to right represent different combinations of filters as indicated in the plots. It can be observed that as the signal-to-noise ratio increases and the filter combinations become redder, the accuracy of the technique represented by the tightness of the data point in the y-axes improves correspondingly. Cyan error bars indicate the standard deviations of calculated $A_{V,output} - A_{V,input}$.

## 5 SUMMARY

In this paper, we investigate the feasibility of studying inter-arm molecular clouds in nearby galaxies through optical extinctions utilizing multi-band optical/IR observations from space-based telescopes.

We first derive dust extinction maps of 17 inter-arm massive molecular clouds in M31 using the PHAT data (Dalcanton et al. 2012) with the CMD-based method proposed by Dalcanton et al. (2015). The majority of the inter-arm massive molecular clouds show an optical $A_V \sim 1$ mag. Only three of the inter-arm massive molecular clouds with $A_V \gtrsim 1.6$ mag exhibit an $A_V$ excess compared to their surroundings, while massive molecular clouds with lower $A_V$ do not show any obvious $A_V$ excess. The derived $A_{V,CMD}$ aligns more closely with measurements using *JCMT* observations (Jiao et al. in prep.). We therefore assert that the CMD method is effective for studying inter-arm massive molecular clouds with high optical extinctions ($A_V > 1.6$) in M31. This assertion carries significant implications as we can reasonably infer that a substantial portion of molecular clouds in nearby galaxies can be detected using this method. We therefore claim that the CMD method is an effective way for studying inter-arm massive molecular clouds with high optical extinctions ($A_V > 1.6$) in M31, and this holds significant implications because we can reasonably infer that we can detect a considerable portion of molecular clouds in nearby galaxies using this method.

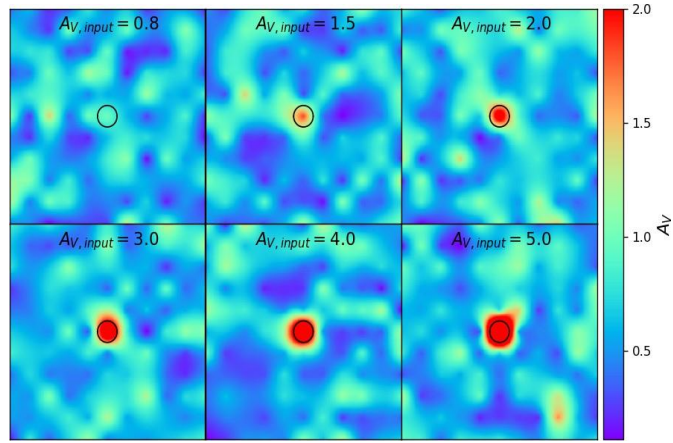

Fig. 9: The extinction distribution map of massive molecular clouds with different extinction values we apply, where the black circle represents the location of the massive molecular cloud. The text at the top of each panel indicates the calculated extinction value of the cloud.

We then test our method using a *CSST* mock RGB star catalog with different input $A_V$ values. We show that the derived $A_V$ values using $z$ and $y$ photometries align more closely with the input values, underscoring the applicability of this method with the *CSST* data. Furthermore, we derive optical extinction maps around molecular clouds with different $A_V$ values and find that molecular clouds with high optical extinctions ($A_V > 1.6$) can be easily identified in inter-arm regions. This highlights the potential of this technique, especially when combined with the extensive survey capabilities of *CSST*, in the systematic identification of inter-arm molecular clouds with high optical extinctions.

This work is supported by National Science Foundation of China Nos. 11988101, 12373012, and 12041302. This work is also supported by CMS-CSST-2021-A08 and CMS-CSST-2021-B02. The *CSST* simulated data were reduced by the *CSST* scientific data processing and analysis system of the China Manned Space Project. Y.Y. acknowledge the support from NSFC with Grant No.12203064. This research is based on observations made with the NASA/ESA Hubble Space Telescope obtained from the Space Telescope Science Institute, which is operated by the Association of Universities for Research in Astronomy, Inc., under NASA contract NAS 5–26555.

### References

Bok, B. J. 1956, AJ, 61, 309

Bok, B. J., McCuskey, S. W., Cherry, B., & Shapley, H. 1937, Annals of Harvard College Observatory, 105, 327

Cardelli, J. A., Clayton, G. C., & Mathis, J. S. 1989, ApJ, 345, 245 Chen, Y., Girardi, L., Fu, X., et al. 2019, A&A, 632, A105

Chevance, M., Krumholz, M. R., McLeod, A. F., et al. 2023, in Astronomical Society of the Pacific Conference Series, 534, 1

Colombo, D., Hughes, A., Schinnerer, E., et al. 2014, ApJ, 784, 3

Dalcanton, J. J., Williams, B. F., Lang, D., et al. 2012, ApJS, 200, 18

Dalcanton, J. J., Fouesneau, M., Hogg, D. W., et al. 2015, ApJ, 814, 3

Dame, T. M., Ungerechts, H., Cohen, R. S., et al. 1987, ApJ, 322, 706

Dickman, R. L. 1978, AJ, 83, 363

Draine, B. T., Aniano, G., Krause, O., et al. 2014, ApJ, 780, 172

Foreman-Mackey, D., Hogg, D. W., Lang, D., & Goodman, J. 2013, PASP, 125, 306

Gao, Y., & Solomon, P. M. 2004, ApJ, 606, 271

Habing, H. J., Miley, G., Young, E., et al. 1984, ApJ, 278, L59

Hu, T., & Peng, Q.-H. 2014, RAA, 14, 869-874

Jeffreson, S. M. R., Sun, J., & Wilson, C. D. 2022, MNRAS, 515, 1663

Jiao, S., Lin, Y., Shui, X., et al. 2022, Science China Physics, Mechanics, and Astronomy, 65, 9

Kirk, J. M., Gear, W. K., Fritz, J., et al. 2015, ApJ, 798, 58

Koda, J., Scoville, N., Sawada, T., et al. 2009, ApJ, 700, L132

Lada, C. J., Lada, E. A., Clemens, D. P., & Bally, J. 1994, ApJ, 429, 694

Lee, J. C., Whitmore, B. C., Thilker, D. A., et al. 2022, ApJS, 258, 10

Lee, J. C., Sandstrom, K. M., Leroy, A. K., et al. 2023, ApJ, 944, L17

Leroy, A. K., Walter, F., Brinks, E., et al. 2008, AJ, 136, 2782

Leroy, A. K., Walter, F., Sandstrom, K., et al. 2013, AJ, 146, 19

Leroy, A. K., Schinnerer, E., Hughes, A., et al. 2021, ApJS, 157, 43

Lindberg, C. W., Murray, C. E., Dalcanton, J. J., Peek, J. E. G., & Gordon, K. D. 2024, ApJ, 963, 58

Lombardi, M., & Alves, J. 2001, A&A, 377, 1023

Lynds, B. T. 1962, ApJS, 7, 1

Rice, W. 1993, AJ, 105, 67

Schlafly, E. F., & Finkbeiner, D. P. 2011, ApJ, 737, 103

Solomon, P. M., Rivolo, A. R., Barrett, J., & Yahil, A. 1987, ApJ, 319, 730

Stark, A. A., & Lee, Y. 2005, ApJ, 619, L159

Wang, Y., Gao, J., & Ren, Y. 2022, ApJS, 259, 12

Williams, B. F., Lang, D., Dalcanton, J. J., et al. 2014, ApJS, ApJS, 215, 9

Williams, B. F., Bell, E. F., Boyer, M. L., et al. 2021a, HST Proposal. Cycle 29, ID. #16778, 16778

Williams, B. F., Durbin, M. J., Dalcanton, J. J., et al. 2021b, ApJS, 253, 53

Zhan, H. 2011, SCIENTIA SINICA Physica, Mechanica Astronomica, 41, 1441

Zhan, H. 2021, Chinese Science Bulletin, 66, 1290

Zhang, X., Cao, L., & Meng, X. 2019, Ap&SS, 364, 9